\documentclass[fleqn,usenatbib]{mnras}
\usepackage{newtxtext,newtxmath}
\usepackage[T1]{fontenc}
\DeclareRobustCommand{\VAN}[3]{#2}
\let\VANthebibliography\thebibliography
\def\thebibliography{\DeclareRobustCommand{\VAN}[3]{##3}\VANthebibliography}
\usepackage{graphicx}	
\usepackage{amsmath}	
\usepackage{ulem}

\title[Can EDE be probed by high-z galaxy abundance?]{Can Early Dark Energy be Probed by the High-Redshift Galaxy Abundance?}

\author[Liu W. et al.]{
Liu, Weiyang,$^{1,2}$\thanks{E-mail: \href{mailto: wyliu@bao.ac.cn}{wyliu@bao.ac.cn} (NAOC)}
Zhan, Hu,$^{1,3}$
Gong, Yan,$^{1,2}$
and Wang, Xin$^{1,2,4}$
\\
$^{1}$Key Laboratory of Space Astronomy and Technology, National Astronomical Observatories, Chinese Academy of Sciences, Beijing, 100101, China P.R. \\
$^{2}$School of Astronomy and Space Science, University of Chinese Academy of Sciences, Beijing, 100049, China P.R. \\
$^{3}$The Kavli Institute for Astronomy and Astrophysics, Peking University, Beijing, 100871, China P.R. \\
$^{4}$Institute for Frontiers in Astronomy and Astrophysics, Beijing Normal University, Beijing, 102206, China P.R.}

\date{Accepted XXX. Received YYY; in original form ZZZ}

\pubyear{2024}

\begin{document}
\label{firstpage}
\pagerange{\pageref{firstpage}--\pageref{lastpage}}
\maketitle

\begin{abstract}
The analysis of the Cosmic Microwave Background (CMB) data acquired by the Atacama Cosmology Telescope (ACT) and the large-scale ($\ell\lesssim1300$) Planck Telescope show a preference for the Early Dark Energy (EDE) theory, which was set to alleviate the Hubble tension of the $\Lambda$ Cold Dark Matter ($\Lambda$CDM) model by decreasing the sound horizon $r_{s}$, and gives $H_{0} \approx 72$ km s$^{-1}$ Mpc$^{-1}$. However, the EDE model is commonly questioned for exacerbating the $\sigma_8$ tension on top of the $\Lambda$CDM model, and its lack of preference from the late-time matter power spectrum observations, e.g., Baryon Oscillation Spectroscopic Survey (BOSS). In light of the current obscurities, we inspect if the high redshift galaxy abundance, i.e., Stellar Mass Function/Density (SMF/SMD) and Luminosity Function (LF), can independently probe the EDE model. Our result shows that, compared to $\Lambda$CDM, the EDE model prediction at $z>10$ displays better consistency with the unexpectedly high results observed by the James Webb Space Telescope (JWST). At lower redshift, the EDE model only fits the most luminous/massive end, with the majority of the data presenting better consistency with $\Lambda$CDM, implying that adding an extra luminosity/mass-sensitive suppression mechanism of the galaxy formation is required for EDE to explain all data around $z\sim7-10$.
\end{abstract}



\section{Introduction}

The discovery of the accelerating expansion of the Universe and the resulting cosmological constant dark energy (\cite{1998AJ....116.1009R,1999ApJ...517..565P} ) led to the transition of the cosmological model from Einstein de Sitter ($\Omega_{m}=1$) to $\Lambda$CDM Universe. This was later recognised as the standard model of cosmology for it was endorsed by numerous follow-up observations. However, despite the success, the last decade witnessed growing doubts regarding this model, due to the appearance of the (third) Hubble tension that implies the values of the Hubble constant measured from the early and late Universe are inherently inconsistent. For instance, when assuming $\Lambda$CDM, \cite{2020A&A...641A...6P} derived $H_{0}=67.36\pm0.54$ km s$^{-1}$ Mpc$^{-1}$ from CMB, while the direct local observational results are significantly higher (\cite{2021CQGra..38o3001D}). For example, Supernovae and $H_{0}$ for the Equation of State of dark energy
(SH0ES, \cite{2022ApJ...934L...7R}) showed that $H_{0}=73.30\pm1.04$ km s$^{-1}$ Mpc$^{-1}$. Although many efforts have been made to check whether the inconsistency between the local direct and early indirect measurement of $H_{0}$ is caused by some statistical flukes or overlooked observational errors, the consensus still has not been reached as yet. Instead, the rising concern is that the modification of the cosmological model itself is required to alleviate the tension.

On top of the Hubble tension, the $\Lambda$CDM model is also challenged by the anomaly on the matter clustering, dubbed as the $S_{8}$ tension. The factor $S_{8}\equiv\sigma_{8}\sqrt{\Omega_{m}/0.3}$ measures the amplitude of the matter clustering and is closely related to the growth rate of the large-scale structure via $f\sigma_{8}$, where $f=[\Omega_{m}(z)]^{0.55}$. The observed result of this factor from the Weak Lensing (WL), e.g., Kilo-Degree Survey (KiDS), is $S_{8}\approx0.759$ (\cite{2021A&A...645A.104A}), about $2-3 \sigma$ lower than the expected value from CMB that gives $S_{8}\approx0.834$ (\cite{2020A&A...641A...8P}). Likewise, there are two possible approaches towards the alleviation of the $S_{8}$ tension, the observational/statistical errors, and the alteration of the model, the latter of which may imply anomalies on the matter power spectrum (e.g., \cite{2023PhRvD.107l3538P}) or galaxy formation process that lead to the consequential redshift-dependent $S_{8}$ observational value. The consensus, however, has not yet been reached either. More information about the aforementioned tensions can be found in \cite{2022JHEAp..34...49A}.

\cite{2022PhR...984....1S} summarised and compared the most common 17 theories proposed to resolve the Hubble tension and among them, the EDE theory is one of the few that carries both theoretical capability and observational evidence. The basic idea of EDE is to introduce a scalar field that behaves like dark energy before the recombination and dilutes faster than the radiation such that the post-recombination evolution remains to be $\Lambda$CDM. This extra component decreases the physical sound horizon $r_s$ by slightly increasing the cooling rate of the Universe. As a result, the corresponding $H_{0}$ inferred from CMB is increased accordingly.

The most direct observational evidence of EDE by far comes from \cite{2021PhRvD.104l3550P, 2022PhRvD.105l3536H}, in which they analysed the CMB temperature power spectrum data from the Wilkinson Microwave Anisotropy Probe (WMAP), ACT and Planck Telescope. The results show that although EDE is not favoured by the full-scale Planck data, the preference of the model is shown in WMAP+ACT which gives $H_{0}=73.43_{-3.4}^{+2.6}$ km s$^{-1}$ Mpc$^{-1}$. As Planck and ACT mainly differ on the small scale, $\ell\gtrsim1000$, \cite{2022PhRvD.106d3526S} further indicated that the Planck data truncated to $\ell<1300$ also shows a preference for EDE. On the other hand, the observational result of the late-time matter power spectrum made by BOSS, e.g., \cite{2020PhRvD.102d3507H}, does not reveal clear signs of EDE. However, other studies such as \cite{2023PhRvD.107f3505S,2022ApJ...929L..16H} argued that the conclusion drawn from BOSS power spectrum could be affected by the other non-cosmological factors, e.g., volume effects or the normalisation of the window function. With the correction of the latter, \cite{2023PhRvD.108d3513H} found that Planck+BOSS returns $H_{0}=70.57\pm1.36$ km s$^{-1}$ Mpc$^{-1}$, and the tension against SH0ES is reduced to $1.4\sigma$ in the frequentist profile likelihood analysis (see also \cite{2021JCAP...11..031B} for more information about the correction of the window function). 
More studies attempting to search for the trace of EDE can be found in \cite{2022PhRvD.105j3514J,2020PhRvD.102l3523L,2023arXiv230300746G,2022PhRvD.106j3522G,2023MNRAS.520.3688R,2022arXiv220812992S,2021PhRvD.103f3502M}; \cite{2021PhRvD.104f3524V}, in which various probes (e.g., weak lensing, Lyman-$\alpha$ forest, and early Integrated Sachs-Wolfe effect) and alterations of the model (e.g., massive neutrinos) were applied. 
Generally speaking, as is pointed out by \cite{2023Univ....9..393V}, the full resolution of the Hubble tension may rely on the combination of the new physics of the early, the late, and the local Universe, and it is still far from concluding the debate over the observational evidence of EDE at this stage. In light of the ambiguity of the current observational results, it is indispensable to verify the model with another probe and break the stalemate on the small-scale and post-recombination epoch.

In this paper, we aim to study if the abundance of the galaxies at high redshift is capable of probing the EDE. This idea was first proposed by \cite{2021MNRAS.504..769K}, in which they calculated the Halo Mass Function (HMF) in the EDE scenario and found that there are more halos formed than in the $\Lambda$CDM case. Interestingly, this phenomenon is more prominent at higher redshift. Hence, the galaxy abundance, such as HMF, satisfies what is required to study the EDE model in both ways: 1. The observation of the galaxy abundance can be conducted on small-scale surveys, which makes it possible to extract information from the galaxy distribution at ultra-high redshift. Note that the typical redshift of the BOSS matter power spectrum is below $z\sim1$ (see e.g., \cite{2023PhRvD.107f3505S,2020PhRvD.102d3507H}); 2. The expectation of a predominant outnumbered halo number density at ultra-high redshift in the EDE model could make it distinguishable from $\Lambda$CDM. Thanks to the deep field observation archives from e.g., Hubble Space Telescope (HST) and Spitzer Infrared Array Camera (\textit{Spitzer}/IRAC), as well as the advent of the James Webb Space Telescope data, the detection of the EDE on the high redshift galaxy abundance is now ready. 

An unexpected feature observed by JWST is that it appears to suggest that the stellar mass density of the massive galaxies at redshift $z>7$ is considerably higher than the extrapolation from the previous studies (\cite{2023Natur.616..266L}). Specifically, \cite{2023NatAs...7..731B} argued that these massive galaxies lie on the upper limit of what is allowed by physics, which requires more than 50 per cent of the baryons to be converted into the stellar mass of these objects. Although there are also other speculations regarding the adjustment of the Initial Mass Function (IMF), e.g., \cite{2023ApJ...951L..40S}, or the uncertainty on the photometric redshift (\cite{2023arXiv230315431A}), this surprising discovery may imply that the current galaxy formation theory or the $\Lambda$CDM itself require further contemplation. As we shall see, the predicted stellar mass density under the EDE model displays better consistency with these unexpectedly high results.

On top of all of the concerns above, we also notice that the change in cosmology itself can simultaneously affect the estimation of the stellar mass through the change of the galaxy formation history. Alternatively, one might stick with the luminosity of the galaxies to avoid the uncertainty of the stellar mass estimation, which involves the composite of cosmology and various baryonic interaction processes. However, the observational result of the luminosity depends on the knowledge of the distance modulus and the dust attenuation (e.g., \cite{1999ApJ...521...64M}) that may behave differently at the ultra-high redshifts. Bearing all in mind, we will show that the luminosity is not noticeably affected by EDE and leave the full-scale estimation of the stellar mass under EDE for future work since it is beyond the topic of this paper.

The structure of this paper is as follows. In Section \ref{sec:ede}, we introduce the theory of EDE. Then we present the way to predict the high redshift galaxy abundance in Section \ref{sec:abundance}. The comparison of our predictions with the observational data is described in Section \ref{sec:results}. Finally, we conclude this paper in Section \ref{sec:conclusions}.

\section{Early Dark Energy Model}
\label{sec:ede}

The angular size of the sound horizon $\theta_{s}\equiv r_{s}/D_{A}$ at the epoch of recombination is precisely measured by the CMB observations. Here $r_{s}$ and $D_{A}$ are the physical size of the sound horizon and the distance towards the surface of the last scattering, respectively. Starting from a fixed $\theta_{s}$ and a given cosmological model as the preconditions, we can derive the corresponding $H_{0}$ as follows (\cite{2023ARNPS..73..153K}),
\begin{align}
    H_{0} = \sqrt{3}H_{ls}\theta_{s} \frac{\int_{0}^{\infty} dz [\rho(z)/\rho_{0}]^{-1/2}}{\int_{z_{ls}}^{\infty} dz[\rho(z)/\rho(z_{ls})]^{-1/2} (1+R)^{-1/2}}, 
    \label{eq:h_cmb}
\end{align}
where $H_{ls}$ is the Hubble parameter at the last scattering, $z_{ls}\simeq1080$. $\rho(z)$ and $\rho_{0}$ represent the total energy density at redshift $z$ and present, respectively. $R=(3/4)(\omega_{b}/\omega_{\gamma})/(1+z)$ is the density ratio of the baryons to photons at $z$. By decreasing the integral of the denominator, i.e., increasing the energy density $\rho(z)$ before recombination, the inferred $H_{0}$ can be elevated accordingly, and the corresponding physical size of the sound horizon
\begin{align}
    r_{s}=\int_{z_{ls}}^{\infty} \frac{c_{s}(z)dz}{H(z)}
    \label{eq:rs}
\end{align}
is decreased. Here $c_{s}(z)=c[3(1+R)]^{-1/2}$ is the sound speed of the baryon-photon fluid. The collection of the methods following this type of mechanism is normally dubbed as the \textit{Early-time} solution, of which a representative example is the EDE model we apply in this paper. 

The EDE model by far is the collective name of an ad hoc postulation with various alterations of dynamics. In this paper, we focus on the most studied \textit{Axion-like} EDE model (\cite{2016PhRvD..94j3523K,2018PhRvD..98h3525P,2019PhRvL.122v1301P,2023ARNPS..73..153K}), which is shown to be related to string theory (\cite{2022arXiv220900011M,2023JHEP...06..052C}). The EDE model introduces an oscillating axion scalar field $\phi$ with a potential,
\begin{align}
    V_n(\phi) = m^{2}f^{2}[1 - \cos{(\theta)}]^{n}
\end{align}
to accelerate the cooling rate of the Universe before the recombination takes place. Here $m$ denotes the effective mass of the axion particle, $f$ represents the decay constant, and $\theta\equiv\phi/f$ is a re-normalisation factor so that $-\pi\leq\theta\leq\pi$. The decay rate of the axion field is determined by the index $n$ via the equation of state $w_{n}=(n-1)/(n+1)$. When $n=1$, it behaves like the normal dark matter axion field. However, if $n>2$ the field decays faster than radiation so that the post-recombination epoch remains to be the $\Lambda$CDM-like Universe. The EDE field is initially frozen at a fixed value until the critical redshift $z_{c}$, after which it starts to oscillate and decay. The fractional energy density of EDE at this point is denoted by $f_{\text{EDE}}(z_{c})$. The preliminary attempts to constrain the EDE parameters mainly focus on the CMB observation, e.g., \cite{2018PhRvD..98h3525P,2019PhRvL.122v1301P}. Specifically, as is indicated by \cite{2021PhRvD.104l3550P}, if $n=3$, the best-fit EDE parameters reconstructed from the WMAP+ACT+BAO+Pantheon data are $f_{\text{EDE}}=0.158_{-0.094}^{+0.051}$ and $\log_{10}(z_{c})=3.326_{-0.093}^{+0.2}$, which give $H_{0}=73.43_{-3.4}^{+2.6}$ km s$^{-1}$ Mpc$^{-1}$ and significantly reduce the Hubble tension against SH0ES.

There has been a growing number of studies attempting to search for more evidence of EDE from CMB. \cite{2022PhRvD.106d3526S} took the South Pole Telescope (SPT) data into account and found that ACT+SPT+Planck polarisation gives a similar conclusion that favours the EDE model. They also found that the Planck temperature power spectrum shows a preference for EDE if the small-scale data ($\ell>1300$) are discarded. More works, e.g., \cite{2022PhRvD.105j3514J, 2022PhRvD.105l3536H}, also found pro-EDE evidence from the observations of CMB.

However, applying the EDE-like mechanism to resolve the Hubble tension comes with the cost of an exacerbated $S_{8}$ tension since it results in a denser Universe. Generally speaking, the locally observed values of $S_{8}$ from e.g., Weak Lensing, are lower than those inferred from $\Lambda$CDM. For instance, \cite{2021A&A...645A.104A} analysed the data from KiDS and obtained $S_{8}\simeq0.759$, while the result from \cite{2020A&A...641A...8P} is about $2-3 \sigma$ higher and reaches $S_{8}\simeq0.834$. In comparison, the result from \cite{2021PhRvD.104l3550P} showed that $S_{8}\simeq0.862$ in the EDE model, which is slightly higher than the $\Lambda$CDM case. As a consequence, most of the post-recombination observations do not show any preferences on EDE over $\Lambda$CDM. For instance, \cite{2023arXiv230300746G} attempted to use Lyman-$\alpha$ forest to constrain the EDE parameters and found that $f_{\text{EDE}}<0.08$ with $H_{0}=67.9_{-0.4}^{+0.4}$ km s$^{-1}$ Mpc$^{-1}$, $>4\sigma$ away from the SH0ES result (\cite{2022ApJ...934L...7R}). On top of that, the signal of EDE does not emerge in the late-time large-scale structure (LSS) observations, either. \cite{2020PhRvD.102d3507H} reanalysed the EDE scenario with the LSS data, including CMB lensing, Baryon Acoustic Oscillation (BAO), Redshift Distortion (RSD), Dark Energy Survey Year 1 (DES-Y1), Hubble Source Catalog (HSC) and KiDS, and found no clear signal of EDE. 
Although \cite{2023PhRvD.107f3505S} reassessed the BOSS data by applying a new normalisation method of the window function and slightly ameliorated its disfavour over EDE, the challenge of $S_{8}$ tension remains intact for the increment of the matter density is an inevitable consequence of EDE. Similar to $\Lambda$CDM, the resolution of it depends on whether there should be a temporal mechanism that results in the evolution of the matter clustering over redshift.

Briefly speaking, despite the ``spark of hope'' to resolve the Hubble tension lit by the EDE theory, the mechanism behind it is still under challenges from the \textit{small-scale} Planck temperature power spectrum and the \textit{low-redshift} observations. The detailed reviews of the current dilemma of EDE in 2023 can be found in \cite{2023arXiv231019899M,2023arXiv230209032P}. Be that as it may, one notices that the current analyses focus on the recombination and late-time Universe while omitting the observations of the post-recombination high redshifts, e.g., the Cosmic Dawn and the reionisation. We argue that the study of EDE in these epochs can complement what is needed for resolving the current stalemate: 1. It provides an alternative way to study the small-scale problem of EDE; 2. It is relatively less complicated by the late-time non-linear process of the cosmic structure/galaxy formation history which is one of the speculations of the reason for the $S_{8}$ tension. A new probe from the realm of the small-scale and post-recombination high redshift is yearned for the verification of EDE. As we shall see below, the galaxy abundance might indeed bear the probability to deliver the just cause for it.






\section{Galaxy Abundance in the EDE Scenario}
\label{sec:abundance}

The basic idea of the galaxy formation theory from the modern perspective is that the dark matter halo distribution and merger history form the skeleton of the galaxy distribution. Then, the baryons gravitationally bound by these systems start to form stars, during which various kinds of feedback (such as stellar and AGN feedback) take place and suppress or quench the further star formation process. The basic empirical approach towards galaxy formation theory is the \textit{Abundance matching}, which implies that the most massive galaxies reside in the most massive halos (\cite{2018ARA&A..56..435W}). Consequently, the predicted abundance of the galaxies, normally expressed in terms of the stellar mass function or luminosity function, depends on the halo mass function and the scaling relation between the halo mass $M_{h}$, stellar mass $M_{\ast}$, and luminosity at the ultraviolet band $L_{\text{UV}}$ (or the absolute magnitude $M_{\text{UV}}$). In Section \ref{subsec:hmf}, we compare the halo mass functions in the EDE and $\Lambda$CDM models. In particular, We adopt the EDE (CMB-only, ACT+SPT+truncated Planck) and $\Lambda$CDM parameters inferred by \cite{2022PhRvD.106d3526S} and \cite{2020A&A...641A...6P}, respectively. The exact values of these parameters are listed in Table \ref{tab:ede_params}. Specifically, the combination of pure CMB data returns a high EDE fraction ($f_{c}=0.163$), which results in an asymptotic scale-independent Harrison-Zel'dovich primordial power spectrum ($n_{s}=1.001$, also see \cite{2023MNRAS.526L..63P} for a more detailed analysis on how the primordial power spectrum affects the number density of the galaxies) and a higher matter clustering ($\sigma_{8}=0.8446$). In comparison, The Planck+BOSS data (\cite{2022ApJ...929L..16H}) gives a partial alleviation of the Hubble tension ($H_{0}=70$ km s$^{-1}$ Mpc$^{-1}$), and a smaller EDE parameter set $(f_{c}, n_{s}, \sigma_{8})=(0.07, 0.9704, 0.825)$. In this work, we apply the CMB-only result for it is independent of the observations at lower redshift, and provides higher $H_{0}$ and EDE fraction which could return an upper limit of the galaxy abundance that is allowed by the current EDE best fit. Then, we apply the empirical scaling relations to derive the corresponding stellar mass functions in Section \ref{subsec:smf} and luminosity functions in Section \ref{subsec:lf}.

\begin{table}
    \centering
    \begin{tabular}{lcc}
        \hline
        \hline
        Parameter       & EDE       & $\Lambda$CDM  \\
        \hline
        $h$             & 0.7420    & 0.6736        \\
        $\omega_{c}$    & 0.1356    & 0.1200        \\
        $n_{s}$         & 1.0010    & 0.9649        \\
        $\sigma_{8}$    & 0.8446    & 0.8111        \\
        $n$             & 3         &               \\
        $\lg({z_{c}})$  & 3.526     &               \\
        $f_{c}$         & 0.163     &               \\
        \hline
    \end{tabular}
    \caption{The cosmological parameters of the EDE and $\Lambda$CDM models to generate the matter power spectrum. The values of EDE are the best fit of ACT+SPT+Planck TT650TEEE from \citet{2022PhRvD.106d3526S} and the $\Lambda$CDM case is the result of \citet{2020A&A...641A...6P}.}
    \label{tab:ede_params}
\end{table}

\subsection{Halo Mass Function}
\label{subsec:hmf}

Broadly speaking, the derivation of the halo mass function depends on two aspects: the cosmology that mainly affects the matter power spectrum and the properties of the dark matter that determine the formation and evolution of the halos. In EDE theory, the properties of the dark matter stay unchanged, such that we only need to consider the change of the matter power spectrum for the approximation of the halo mass function in the EDE/$\Lambda$CDM model. In order to stay consistent with previous studies, we make use of the \texttt{AxiCLASS} (\cite{2020PhRvD.101f3523S, 2018PhRvD..98h3525P}), a modified version of the Einstein-Boltzmann code \texttt{CLASS} (\cite{2011JCAP...07..034B}),  to compute the matter power spectra in the scenarios of EDE and $\Lambda$CDM.

\begin{figure*}
    \centering
    \includegraphics[width=\textwidth]{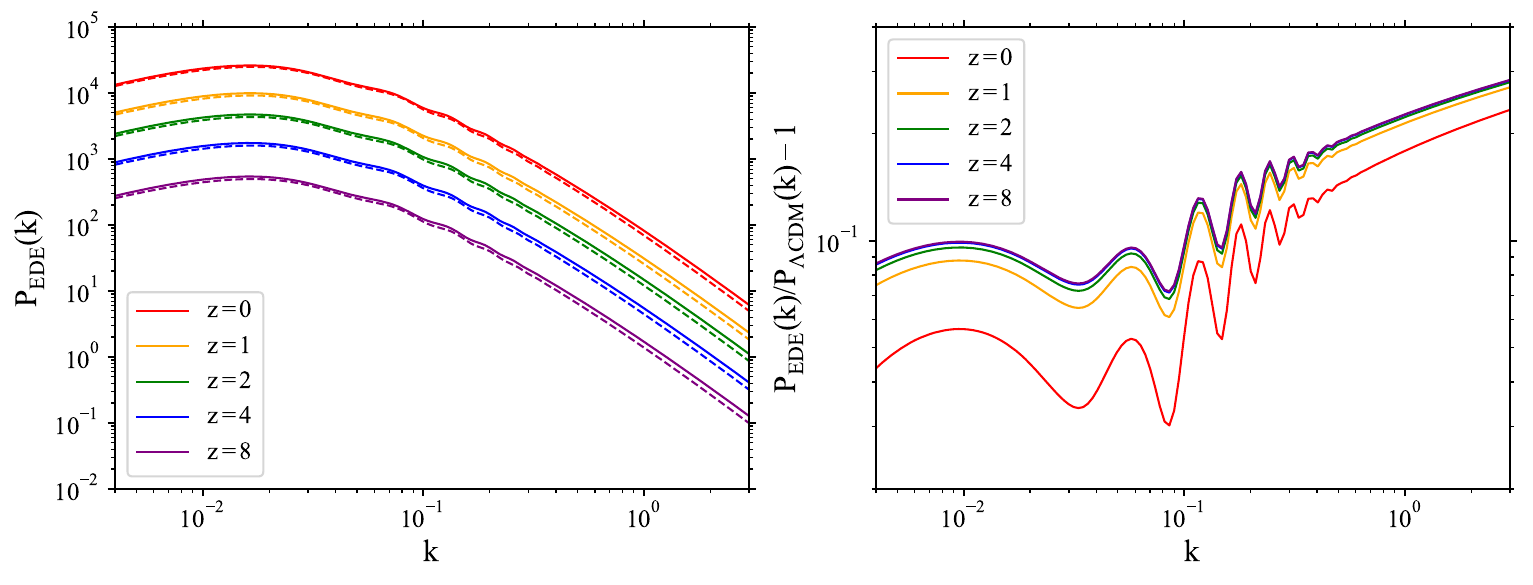}
    \caption{(Left) The matter power spectra of the EDE (solid lines) and $\Lambda$CDM (dashed lines) model. The EDE generates more power at all scales and most prominently at the small scale, due to its larger primordial power spectrum index $n_{s}$; (Right) The ratio of the matter power spectra between the EDE and $\Lambda$CDM model. Their difference decreases as the Universe evolves over redshift.}
    \label{fig:mps}
\end{figure*}

The left side of Figure \ref{fig:mps} shows the matter power spectra of EDE (solid lines) and $\Lambda$CDM model (dashed lines) in 5 different redshifts. It is clear that the EDE generates larger power than $\Lambda$CDM at all scales and, most prominently, at the small scale. This is caused by the slightly increased value of the primordial power spectrum index $n_{s}$ from the best-fit EDE parameters shown in Table \ref{tab:ede_params}. Consequently, the EDE Universe is expected to be denser than the $\Lambda$CDM case (\cite{2021MNRAS.504..769K}). On top of that, if we compare the ratio of the two scenarios which is shown on the right side of Figure \ref{fig:mps}, we see that the difference between them decreases with time, i.e., the Universe becomes more $\Lambda$CDM-like in an EDE Universe as redshift decreases. This phenomenon is significantly more prominent in the halo mass function as we shall see below. Finally, with the matter power spectra prepared, we can calculate the corresponding theoretical halo mass function.

We use the halo mass function approximation from \cite{1999MNRAS.308..119S} to estimate the number of halos in the EDE model \cite[also see][for more information]{2021MNRAS.504..769K, 2016MNRAS.456.2486D}. The halo mass function describes the number density of the halo with mass $M$ at redshift $z$. It is formulated by the smoothed matter power spectrum $\sigma$ as follows,
\begin{align}
    \frac{dn}{dM} = f(\sigma)\left[\frac{\Omega_{m}\rho_{\text{crit}}}{M^{2}}\right] \frac{d\ln{\sigma}}{d\ln{M}}.
    \label{eq:hmf}
\end{align}
Here
\begin{align}
    \sigma^{2}(M, z) \equiv \int \frac{dk}{k} \frac{k^{3}P(k, z)}{2\pi^{2}} W^{2}[kR(M)]
    \label{eq:sigma2}
\end{align}
denotes the extent of the overdensity within the region confined by the Top-Hat window function, $W(x=kR)\equiv3(\sin x - x\cos x)/x^3$. Here $M$ is the mass enclosed inside the radius $R$, $M=(4\pi/3)\rho_{m}R^{3}$. The $P(k,z)$ is the matter power spectrum extrapolated to redshift $z$ as is shown in Figure \ref{fig:mps}. Next, it is commonly accepted to assume that the threshold of the overdensity is $\delta_{\text{crit}}\approx1.68$ at which the density of the overdense region is high enough to form a halo. Define the relative height of the density peak by
\begin{align}
    \nu = \frac{\delta_{\text{crit}}}{\sigma(M, z)},
    \label{eq:nu}
\end{align}
so that the halo mass function can be rewritten as
\begin{align}
    \nu f(\nu) = A\left(1 + \frac{1}{\nu'^{p}} \right) \left(\frac{\nu'}{2\pi} \right) e^{-\nu'/2}, 
    \label{eq:nuf}
\end{align}
where $\nu'=a\nu$ and the simulated coefficients, $(A, a, p)$, determine the normalisation, high mass cut-off, and low mass shape of the function, respectively.

\begin{figure*}
    \centering
    \includegraphics[width=\textwidth]{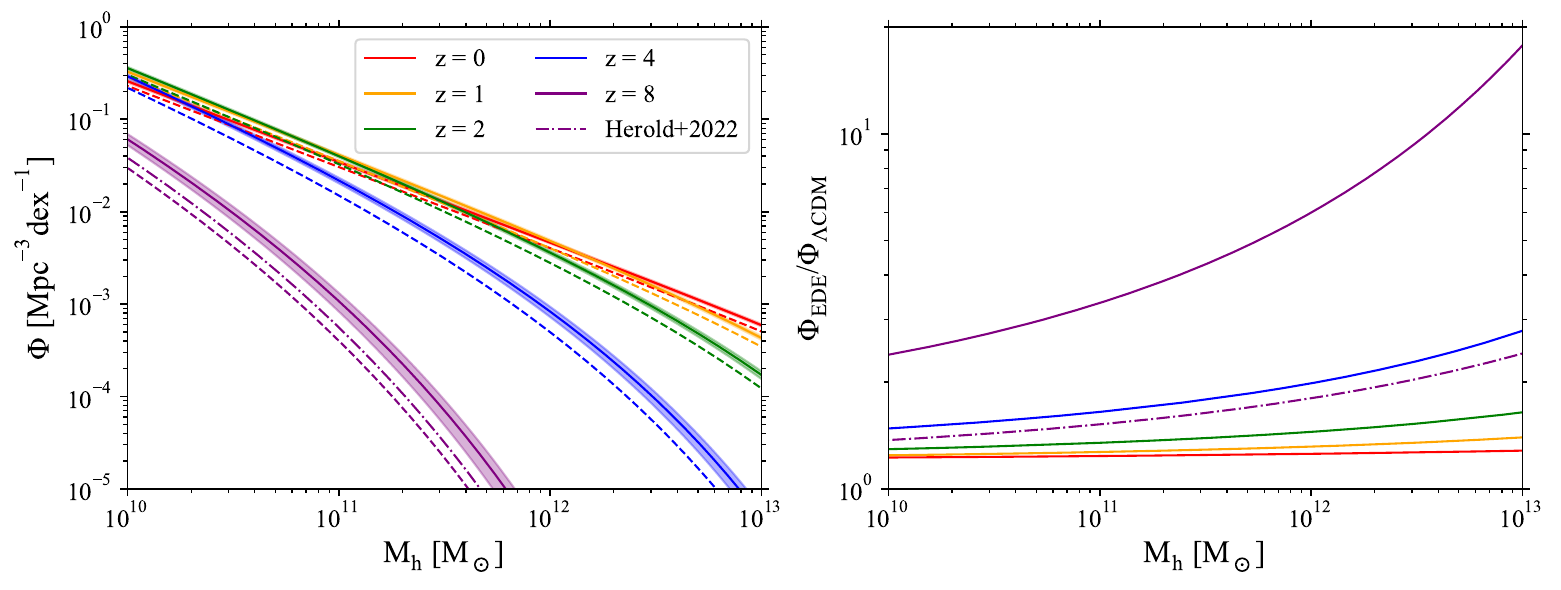}
    \caption{(Left) The halo mass function, $\Phi = dn/d\lg{M_h}$, under the EDE model in different redshifts. The error shade corresponds to the uncertainties on the coefficients $(A, a, p)$ from \citet{2016MNRAS.456.2486D} in Equation \ref{eq:nuf} as well as the uncertainties of the cosmological parameters provided by \citet{2022PhRvD.106d3526S} and \citet{2020A&A...641A...6P}. The uncertainty gets larger at the massive end as the redshift goes higher. We also plot the $\Lambda$CDM cases (dashed lines) for comparison; (Right) The halo mass function ratio between the EDE and $\Lambda$CDM cases. The ratio goes up as the mass and redshift get higher, which implies that EDE can produce more massive galaxies in the early Universe than $\Lambda$CDM. The evolution of their ratio suggests that the observation of the massive galaxies at high redshift should be able to distinguish the two models. We also plot the EDE case at $z=8$ with the parameters fitted by \citet{2022ApJ...929L..16H} on both sides of the figure for comparison (dash-dotted line), which produces a similar number of halos to $\Lambda$CDM due to its smaller fraction of EDE.
    }
    \label{fig:hmf_comp}
\end{figure*}

In practice, we use \texttt{hmf} (\cite{2021A&C....3600487M, 2013A&C.....3...23M})  and adopt the coefficients from \cite{2016MNRAS.456.2486D}, i.e., $(A, a, p) = (0.3295\pm0.0003, 0.7689\pm0.0011, 0.2536\pm0.0026)$ to calculate the halo mass functions for the EDE model. In their work, they found that the halo mass function remains universal for Planck 2013 $\Lambda$CDM Universe and when $\sigma_{8}=0.9$, which is higher than the values of both EDE and Planck 2018 $\Lambda$CDM listed in Table \ref{tab:ede_params}. We therefore argue that their result is a legitimate choice for our purpose, as the elevated value of $\sigma_{8}$ is the main consequence of EDE in the post-recombination epoch. The left side of Figure \ref{fig:hmf_comp} shows the redshift and mass evolution of the halo mass function. Interestingly, the uncertainty gets larger at the massive end as the redshift goes higher. This phenomenon is consistent with the result in \cite{2016MNRAS.456.2486D} since the halo mass function is more sensitive to the normalisation factor $A$ and the high mass cut-off parameter $a$, as $\nu$ gets larger with a higher redshift.

To see the difference between the EDE and $\Lambda$CDM predictions more clearly, we plot the ratio of the halo mass functions in the two models on the right side of Figure \ref{fig:hmf_comp}. As is expected, the EDE model produces more halos in all redshifts. When $z<3$, both cases are almost identical, while the difference of them gets larger at higher redshift. This phenomenon is caused by the fact that the exponential part of Equation \ref{eq:nuf} dominates the halo mass function when $\nu$ is large, which is equivalent to the function becomes more sensitive to $\sigma(M, z)$ as redshift goes higher (\cite{2021MNRAS.504..769K}). Also, we notice that the fraction of EDE can significantly affect the predicted number density of halos. For instance, the dash-dotted line that represents the EDE case at $z=8$ with a smaller $f_{c}$ fitted by Planck+BOSS (\cite{2022ApJ...929L..16H}) is more similar to the $\Lambda$CDM case. Hence, through Equation \ref{eq:sigma2}, the effect of EDE on the power spectrum is revealed via a denser distribution of the dark matter halo. In this case, the feature of the EDE halo mass function is therefore manifested in two aspects: 1. It is consistent with $\Lambda$CDM in the local Universe; 2. It shows an observable deviation from $\Lambda$CDM in the early Universe.

Therefore, bearing the tight bond between the halos and galaxies in mind, according to the abundance matching, we expect that the feature of the outnumbered halo density in the EDE model when $z\geq4$ should also be propagated to the high redshift galaxy abundance, albeit the number density difference of the galaxies is not expected to be as high as the halos due to the baryonic feedback processes of the galaxy formation history and, as was demonstrated in both semi-analytical studies and simulations (\cite{1992MNRAS.256P..43E,2000ApJ...539..517B,2018MNRAS.478..548S}), that most of the low mass halos are unable to form galaxies, both of which suppress or quench the further merging and enrichment of the galaxies.

\subsection{Stellar Mass Function}
\label{subsec:smf}

The mass of the dark matter halo determines the maximum scale that a galactic system can reach. Empirically speaking, the principle of their relationship follows the abundance matching as we mentioned above. However, their relationship is not linear but depends on a composite of various dynamic processes, such as the feedback from star formation or AGN. In particular, the Stellar-to-Halo Mass Ratio (SHMR, or $\epsilon$) at high redshift may differ from the low redshift case, for the stars and AGN need time to form. Notwithstanding the previous simulations being implemented to study the galaxies with high precision, it remains an open question to simulate the formation history of them accurately. Therefore, we apply the empirical SHMR based on known data to establish our prediction of the galaxy abundance, for it contains the authentic information required to reconstruct the relation between halos and galaxies, despite the ambiguity of the specific mechanisms concealed within the black box of the galaxy formation process.

\cite{2021ApJ...922...29S} analysed the stellar mass of the galaxy candidates observed by HST and Spitzer at $z\sim6-10$ and found that the SHMR relation does not strongly depend on redshift. Namely,
\begin{align}
    \frac{M_{\ast}}{M_{h}} = 2N \left[\left(\frac{M_{h}}{M_{c}}\right)^{-\beta} + \left(\frac{M_{h}}{M_{c}}\right)^{\gamma} \right]^{-1}.
    \label{eq:mhmast}
\end{align}
Here $N=0.0297\pm0.0065$ is the normalisation factor, $M_{c}=10^{11.5\pm0.2}$ M$_{\odot}$ is the characteristic halo mass at which the star formation efficiency reaches its maximum. The slope   $\gamma=0.4$ (\cite{2018ApJ...868...92T}) and $\beta=1.35\pm0.26$ denote the shape of the relation at the high-mass and low-mass regimes, respectively.

\begin{figure}
    \centering
    \includegraphics[width=\columnwidth]{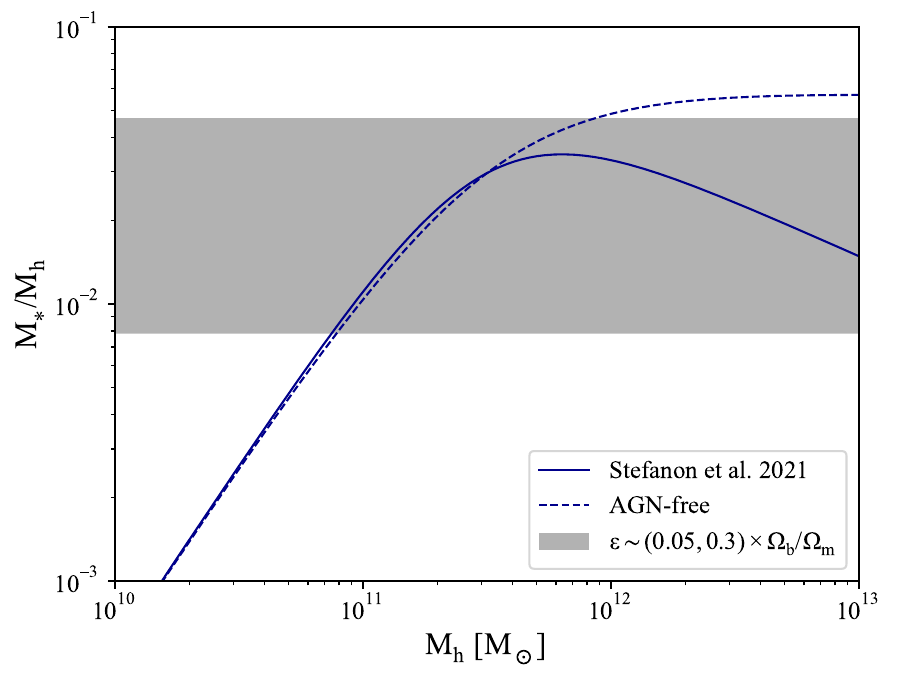}
    \caption{The galaxy formation efficiency proposed by \citet{2021ApJ...922...29S} changes over $M_{h}$ (blue solid line). The grey area marks the constant assumption that 5 to 30 per cent of the baryons are converted into $M_{\ast}$. In particular, the efficiency of the low-mass halos could be as low as $\leq5$ per cent of the total baryonic mass, while the massive halos appear to have a turning point around $M_{h}^{11.5}$ M$_{\odot}$. It is known that the low-mass halos are highly inefficient in forming stars, while the turnover is presumably caused by the feedback from the low-mass AGN or the insufficient merging process between the galaxy progenitors. Alternatively, since the pre-JWST data at $z\geq8$, upon which this relation is built, does not put a strong constraint on the turning point, we also extrapolate the relation to an AGN-free version by letting $\gamma=0.01$ in Equation \ref{eq:mhmast} so that the massive end is flattened (blue dashed line).}
    \label{fig:mhms_ratio}
\end{figure}

The SHMR expression of Equation \ref{eq:mhmast} also indicates that the galaxy formation efficiency is not a constant or a monotonically increasing function, but peaks at around $M_{h}\sim10^{11.5}$ M$_{\odot}$, as is shown in Figure \ref{fig:mhms_ratio} (the blue solid line). In comparison, we also draw the common assumption that $5-30$ per cent of the baryonic mass is converted into stars (the grey area), i.e., $\epsilon=(0.05,0.3)\times\Omega_{b}/\Omega_{m}$. It is clear that the low-mass halos cannot efficiently form galaxies due to their limited capacity to bind the masses contained in them. On the other hand, the turnover of the massive end indicates the potential strong feedback at the high redshift that quenches further enrichment of the stellar mass in these halos. In the low redshift regime, the suppression of the massive end is normally considered the result of the AGN feedback (\cite{2018ARA&A..56..435W}). However, considering only very few numbers of AGN are detected in the galaxies at $z>8$ (e.g., the rare and most distant AGN up-to-date, CEERS\_1019 at $z\approx8.72$, \cite{2023ApJ...953L..29L} and GN-z11 at $z\approx10.603$, \cite{2023arXiv230609142S}), the mechanism behind the turnover above $z\sim8$ might be different from their local counterparts, for AGN also needs time to form. Alternatively, since the SHMR at $z\geq8$ is poorly constrained by the pre-JWST data, we can also speculate that the turnover does not exist at the early stage of the Universe. Therefore, we also extrapolate an AGN-free SHMR by letting $\gamma=0.01$ and remove the turnover (blue dashed line). Instead, if future observations indeed find the turning point, the mechanism behind the suppression at the massive end might be the collective result of the insufficient merging process between the galaxy progenitors within the given time and the early-formed AGN. Nevertheless, considering the redshift-free relation of Equation \ref{eq:mhmast} stands as early as $z=10$, we assume that the potential AGN feedback is not a major concern for applying the low mass end of the SHMR at a higher redshift.

Meanwhile, the redshift-free empirical scaling relation suggests that the early formation process of the galaxies is dominated by the dark matter halos rather than the baryonic compositions which will lead to the evolving feedback mechanisms that alter the formation process over time. According to \cite{2018ApJ...868...92T}, the redshift-free SHMR does not stand below $z\sim4$. Considering the purpose of our work is to investigate the galaxies at high redshift, we therefore cautiously suggest not to extend the usage of this scaling relation for redshift $z<6$, for the late-time galaxies are formed in a much more dusty and complicated environment than their early-time progenitors.

\begin{figure*}
    \centering
    \includegraphics[width=\textwidth]{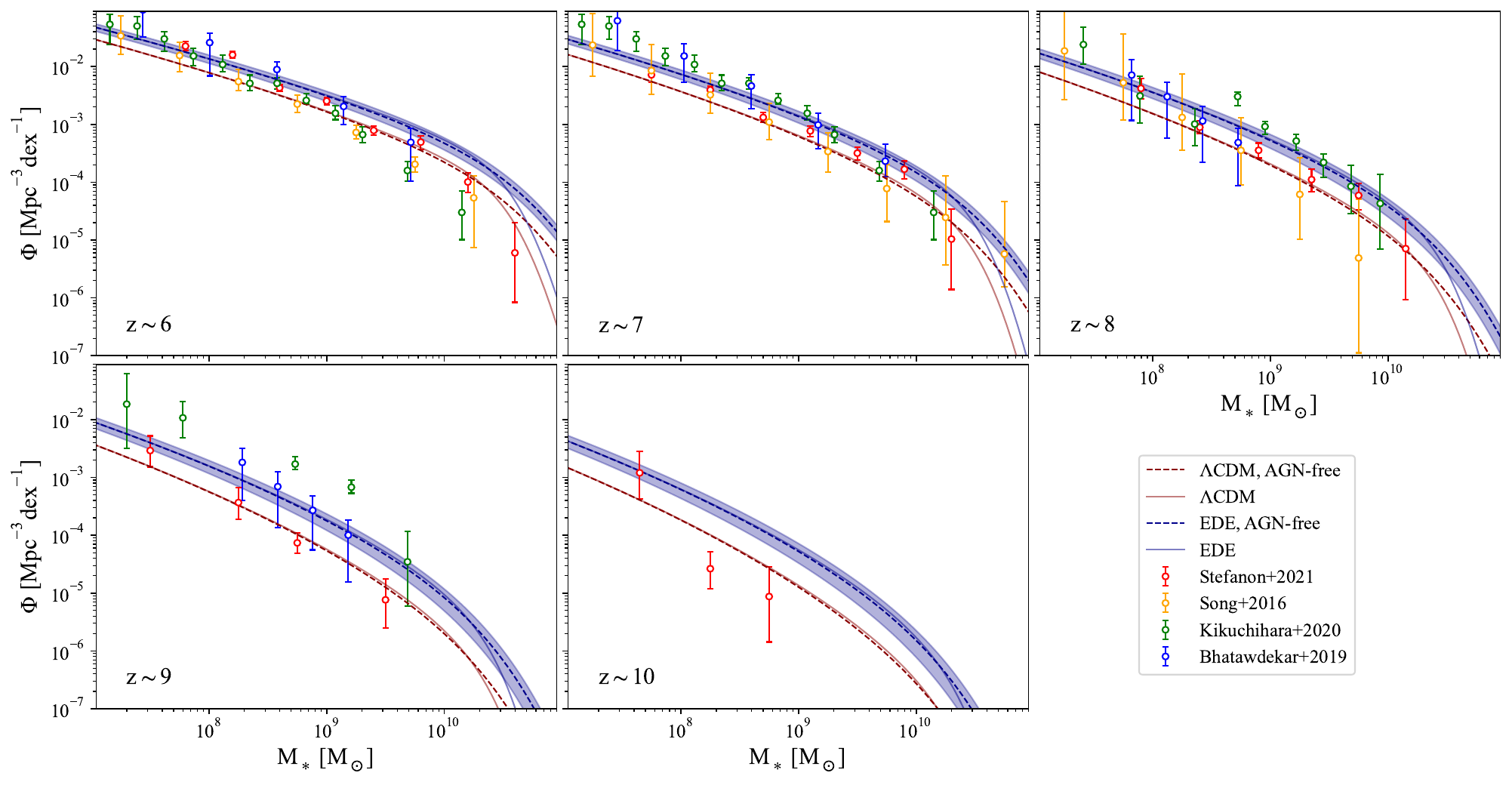}
    \caption{The stellar mass function, $\Phi = dn/d\lg{M_\ast}$, from the EDE (blue) and $\Lambda$CDM (red) model. The AGN-free scaling relation cases (dashed lines) show no significant difference against that from the original, for the modification mainly affects the massive end of the curves. The error shades of the EDE lines are propagated from the uncertainties of the halo mass function (Figure \ref{fig:hmf_comp}) and the $M_{h}-M_{\ast}$ scaling relation parameters $(N, M_{c}, \beta)$ given by \citet{2021ApJ...922...29S}. The data points are from the HST/Spitzer IRAC obtained by \citet{2021ApJ...922...29S,2016ApJ...825....5S,2020ApJ...893...60K,2019MNRAS.486.3805B}. Note that the stellar mass data are not calibrated by the EDE model.}
    \label{fig:smf}
\end{figure*}

Finally, by applying the Jacobian $dM_{h}/dM_{\ast}$ derived from Equation \ref{eq:mhmast} to the halo mass function, Equation \ref{eq:hmf}, we obtain the corresponding stellar mass function as is shown in Figure \ref{fig:smf},
\begin{align}
    \frac{dn}{dM_{\ast}} = \frac{dn}{dM_{h}}\frac{dM_{h}}{dM_{\ast}}.
    \label{eq:jacobian_mhmast}
\end{align}
The blue and red lines in the figure are the EDE and $\Lambda$CDM predictions, respectively. The error shades of the EDE lines are propagated from the uncertainties of the halo mass function (Figure \ref{fig:hmf_comp}) and the SHMR scaling relation parameters $(N, M_{c}, \beta)$. We also plot the HST/Spitzer IRAC data from \cite{2021ApJ...922...29S,2016ApJ...825....5S,2020ApJ...893...60K,2019MNRAS.486.3805B} for comparison. Specifically, the AGN-free scaling relation does not significantly affect the comparison between the theoretical prediction and the data, for the modification only affects the massive end of it. The plots as well as the $\chi^{2}$ (Table \ref{tab:chi2}) show that the predictions from EDE appear to fit better with the existing data. However, it is too soon to draw a definite conclusion as yet, for both models show consistency with specific choices of data sources. As we shall see in the next two sections, the advent of the JWST data may have played an essential role in distinguishing the two models.

The application of the EDE model may also affect the estimation of $M_{\ast}$ due to e.g., the change of the distance measurement, the potentially earlier starting time of the galaxy formation process and a denser cool gas distribution. The distance change may slightly decrease the measured value of the luminosity, as we shall see in Section \ref{subsec:lf}, and thus the estimation of $M_{\ast}$. In contrast, the potential changes in the galaxy formation process itself are expected to lead to an increment in the measurement. In this paper, we apply the vanilla results of the stellar mass data and neglect the possible shift of it under the EDE model, for the scale at which the galaxy formation takes place is way smaller than the context of cosmology, and the properties of dark matter remain unchanged. Nevertheless, a detailed evaluation of the impact that the EDE model may inflict on the stellar mass estimation is required for a more solid argumentation.

\begin{table}
    \centering
    \begin{tabular}{cccc}
        \hline
        \hline
        Redshift    & $\chi^{2}$ ($\Lambda$CDM) & $\chi^{2}$ (EDE)  & $\Delta \chi^{2}$ (EDE-$\Lambda$CDM) \\
        \hline
        \multicolumn{4}{c}{SMF} \\
        \hline
        $6$         & 6.948353                  & 3.568229          & $-$3.380124         \\
        $7$         & 4.602392                  & 2.006349          & $-$2.596043         \\
        $8$         & 0.160697                  & 0.035074          & $-$0.125623         \\
        $9$         & 2.244148                  & 0.771815          & $-$1.472332         \\
        $10$        & 0.001655                  & 0.000395          & $-$0.001260         \\
        \hline
        \multicolumn{4}{c}{LF} \\
        \hline
        $6$         & 0.026143                  & 0.062748          & $+$0.036605         \\
        $7$         & 0.003001                  & 0.015663          & $+$0.012662         \\
        $8$         & 0.001996                  & 0.005233          & $+$0.003237         \\
        $9$         & 2.500958                  & 0.087309          & $-$2.413649         \\
        $10$        & 0.014133                  & 0.008695          & $-$0.005438         \\
        $11$        & 0.016214                  & 0.000958          & $-$0.015257         \\
        $12$        & 7.534370                  & 0.338642          & $-$7.195728         \\
        \hline
    \end{tabular}
    \caption{The $\chi^{2}$ and $\Delta \chi^{2}$ of the Stellar Mass function (Figure \ref{fig:smf}) and Luminosity Function (Figure \ref{fig:lf_data}) in $\Lambda$CDM and EDE model (AGN-free). In SMF, EDE appears to display better consistency with the observed data. For LF, it is the case at $z\geq10$. Note that the result of LF at $z=9$ is dominated by the outliers which only have upper limits of the number density. When they are removed the data favours $\Lambda$CDM.
    }
    \label{tab:chi2}
\end{table}

\subsection{Luminosity Function}
\label{subsec:lf}

In observation, the luminosity function directly counts the number density of the galaxies in terms of the surface brightness with limited hypotheses of cosmology and galaxy formation. The stellar mass function, on the other hand, requires the mass estimation of the galaxies that depends on the theories of both cosmology and the baryonic interactions in the galaxy formation history. Therefore, the detected luminosity function is considered a more direct tool to test the mechanisms of galaxy formation.

The prediction of the luminosity function, however, is unable to be inferred in a model-free way. One of the greatest challenges of predicting the luminosity function from the halo mass function is the difficulty of finding the straightforward scaling relation between the halo mass $M_{h}$ and the luminosity of a galaxy $L_{\ast}$. The $L_{\ast}$ is a direct representative of the properties and distribution of the stars, while the $M_{h}$ does not directly reflect the stellar distribution of the system, i.e., a massive halo may form a galaxy with different compositions of stars according to the age, metallicity and other factors, which correspond to different values of luminosity at a given band. There are attempts to find the luminosity from the halo mass, e.g., \cite{2023MNRAS.521..497M,2022PhRvD.105d3518S}, while some essential assumptions of the galaxies, such as the Star-Formation Rate (SFR), are required to obtain the corresponding $L_{\ast}$, which is equivalent to using the stellar mass $M_{\ast}$ as a proxy to mediate the $M_{h}-L_{\ast}$ relation in between. Following the same logic, we first fix the scaling relation between the stellar mass and luminosity and then derive the corresponding luminosity function as is done in Equation \ref{eq:jacobian_mhmast}, i.e.,
\begin{align}
    \frac{dn}{dM_{\text{UV}}}=\frac{dn}{dM_{h}}\frac{dM_{h}}{dM_{\ast}}\frac{dM_{\ast}}{dM_{\text{UV}}}.
    \label{eq:lf}
\end{align}
Note that we replace the luminosity $L_{\ast}$ by the absolute magnitude at the UV band, $M_{\text{UV}}$, to align with the observations. 

There have been a few number of works discussing the form of the scaling relation between $M_{\text{UV}}$ and $M_{\ast}$, e.g., \cite{2021ApJ...922...29S, 2020ApJ...893...60K, 2019MNRAS.486.3805B, 2014MNRAS.444.2960D}. In their works, the scaling relation is approximated by a linear relation, 
\begin{align}
    M_{\text{UV}} = [\lg({M_{\ast}) - a_{0}}] / a_{1}, 
    \label{eq:muvmast}
\end{align}
where $(a_{0}, a_{1})$ are the slope and intercept of the relation fitted from the pre-JWST data. Here we keep up with the relation provided by \cite{2021ApJ...922...29S} both to keep the consistency and to cover a higher redshift limit which was not reached by the other works. We notice that in their work, $a_{0}$ is not directly given but calculated by $a_{0}=\lg{M_{\ast}}-a_{1}\times-20.5$ where $M_{\ast}$ is the stellar mass corresponds to $M_{\text{UV}}=-20.5$ in each redshift bin (see Table 3 in their work for more information). 

The $M_{\text{UV}}$ converted by Equation \ref{eq:muvmast} is the intrinsic absolute magnitude of a galaxy. In real observations, however, the UV light emitted from the galaxy is absorbed by dust and re-emitted in the Infrared (IR) band. \cite{1999ApJ...521...64M} proposed a simple dust attenuation rule to calibrate the absolute magnitude by using the UV spectral slope $\beta$, which is equivalent to the colour of the galaxy,
\begin{align}
    A_{1600} = 4.43 + 1.99 \beta, 
    \label{eq:dust_att}
\end{align}
where $A_{1600}$ is the dust attenuation factor at 1600 $\mathring{\text{A}}$. Note that the relation is built upon the assumption that the high redshift galaxies are scaled-up starburst galaxies in the local Universe (\cite{1999ApJ...521...64M,2023arXiv231115664S}). Given this, one can obtain the calibration of the dust attenuation in different redshifts with the measurement of $\beta$. 

With the intrinsic luminosity function and the dust attenuation law prepared, we can finally derive a realistic luminosity function prediction in the EDE Universe to compare with the observational results. 






\section{Results}
\label{sec:results}

\begin{figure*}
    \centering
    \includegraphics[width=0.9\textwidth, keepaspectratio]{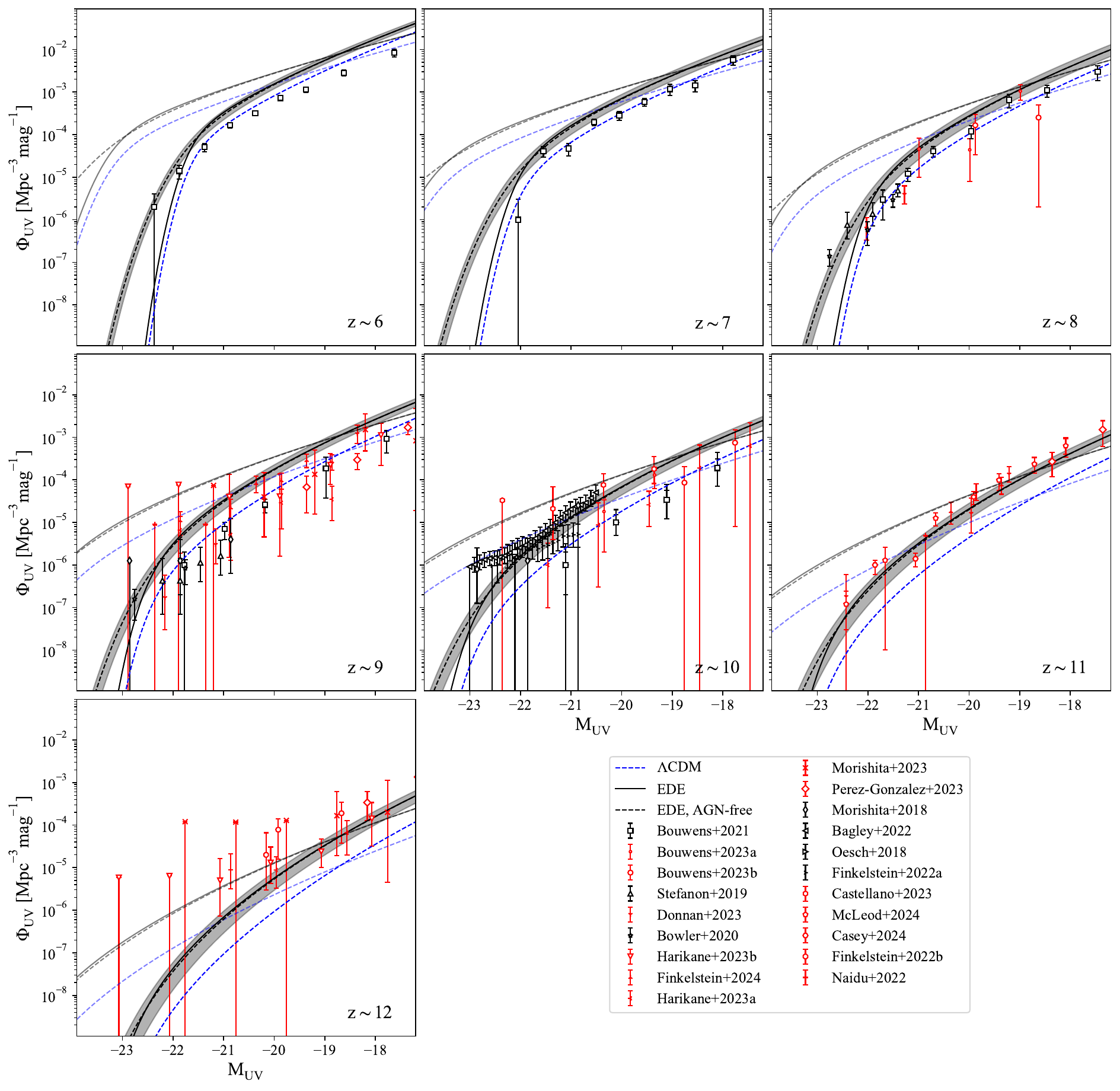}
    \caption{The predicted luminosity functions, $\Phi_{\text{UV}} = dn/dM_{\text{UV}}$, at different redshifts with dust attenuation calibration from \citet{2023MNRAS.520...14C}. The black solid and blue dashed lines correspond to the EDE and $\Lambda$CDM predictions, respectively. We also plot the black dashed lines for the AGN-free scenario (Figure \ref{fig:mhms_ratio}) and the light lines for the dust uncalibrated results. For comparison, we consider both JWST and pre-JWST data from \citet{2023ApJ...948L..14C,2022ApJ...940L..55F,2023arXiv230406658H,2023ApJS..265....5H,2023ApJ...946L..35M,2022ApJ...940L..14N,2023ApJ...951L...1P,2023MNRAS.518.6011D,2023MNRAS.523.1009B,2023MNRAS.523.1036B,2022arXiv220512980B,2021AJ....162...47B,2022ApJ...928...52F,2018ApJ...867..150M,2018ApJ...855..105O,2020MNRAS.493.2059B,2019ApJ...883...99S,2023arXiv231104279F,2024MNRAS.527.5004M,2024ApJ...965...98C}. The black data points are collected from pre-JWST observations, while the red data points are from JWST. We also modify the luminosity distance of the JWST data to make the corresponding $M_{\text{UV}}$ consistent with EDE cosmology. The alteration of the distance brings about $0.15$ right-shift of the $M_{\text{UV}}$ to the fainter end. The error shade collects the uncertainties from both stellar mass function (Figure \ref{fig:smf}) and the $M_{\ast}-M_{\text{UV}}$ scaling relation from \citet{2021ApJ...922...29S}. The results show that when $z\leq9$ the $\Lambda$CDM predictions are consistent with most of the observational results, while the (AGN-free) EDE fits better with the JWST data at $z>10$ and the luminous ends of the lower redshifts. The $z\sim10$ appears to be the boundary of this phenomenon, which implies that a luminosity-sensitive suppression mechanism of the galaxy formation around this era is required if the EDE model is to explain all of the data at $z\lesssim10$.}
    \label{fig:lf_data}
\end{figure*}

The predicted luminosity functions under the EDE and $\Lambda$CDM models as well as the observed results of the pre-JWST and JWST data from \cite{2023ApJ...948L..14C,2022ApJ...940L..55F,2023arXiv230406658H,2023ApJS..265....5H,2023ApJ...946L..35M,2022ApJ...940L..14N,2023ApJ...951L...1P,2023MNRAS.518.6011D,2023MNRAS.523.1009B,2023MNRAS.523.1036B,2022arXiv220512980B,2021AJ....162...47B,2022ApJ...928...52F,2018ApJ...867..150M,2018ApJ...855..105O,2020MNRAS.493.2059B,2019ApJ...883...99S,2023arXiv231104279F,2024MNRAS.527.5004M,2024ApJ...965...98C}, are plotted in Figure \ref{fig:lf_data}. The black data points are from pre-JWST (e.g., HST and Spitzer/IRAC) observations and the red data points are from JWST. We also plot both dust attenuation uncalibrated and calibrated cases of the two models in light and bold lines, respectively. Here we apply the dust attenuation calibration adopted from the JWST $\beta-M_{\text{UV}}$ relation obtained by \cite{2023MNRAS.520...14C}, i.e., $\beta=-0.17M_{\text{UV}}-5.40$. This relation is roughly consistent with the result obtained from the lower redshift while mildly pointing towards the bluer end, implying that the galaxies at high redshift are slightly more active than their local counterpart starbursts. The deviations of the two cases show that the influence of the dust attenuation calibration is less significant at higher redshifts, as the dust also needs time to form. Also, if we assume the SHMR is free from AGN suppression (Figure \ref{fig:mhms_ratio}), the EDE luminosity function will be lifted at the luminous end (black dashed lines). Its effect, however, is limited, as the luminosities of the most observed galaxy candidates are below the turning point extrapolated from the SHMR under the given $M_{\ast}-M_{\text{UV}}$ relation. We also draw the uncertainty of this case that collects the errors from both stellar mass function (Figure \ref{fig:smf}) and the $M_{\ast}-M_{\text{UV}}$ scaling relation given by \cite{2021ApJ...922...29S}.

In addition, since the observational results of $M_{\text{UV}}$ are affected by the background cosmology via the change of the luminosity distance $D_{L}$, we also adjust the data points accordingly. Assuming $(m, M)$ are the apparent and absolute magnitudes of the observed galaxy, they are correlated by
\begin{align}
    m = M + DM + K_{QR}, 
    \label{eq:mM_relation}
\end{align}
where the distance modulus $DM=5\lg[D_{L}/10 \text{ pc}]$ and $K_{QR}$ is the K-correction factor determined by the observed flux and the properties of the detector (\cite{2002astro.ph.10394H}). Therefore, the only factor affected by the background cosmology is $DM$. Apply the EDE model to the calibration of the data and we find that the change of cosmology will slightly shift the corresponding $M_{\text{UV}}$ to the right (dimmer) side by $\sim0.15$ mag. This is not surprising since $D_{L}=(1+z)\times c\int_{0}^{z} 1/H(z') dz'$ decreases with a larger $H_{0}$. Note that in observation it is common to take $H_{0}=70$ km s$^{-1}$ Mpc$^{-1}$.

In the case of $z\sim8$ in Figure \ref{fig:lf_data}, we see that the number density of the galaxies obtained by pre-JWST observations is roughly consistent with the JWST data. As the redshift goes up to $z \sim 9-10$, the pre-JWST results begin to be lower than the JWST cases. The results from \cite{2022arXiv220512980B,2022ApJ...928...52F}, however, are exceptionally higher than other pre-JWST data. Generally speaking, the data from JWST does not show strong evolution among the redshift range $z\sim8-12$. We do not include the data at higher redshift since the future JWST spectroscopic redshift result may considerably affect the current data at $z>12$ due to their limited number of objects and the uncertainties of the photometric redshifts. If we constrain ourselves to the redshift $z\leq12$, the comparison between the data and our prediction clearly shows that the $\Lambda$CDM fits better with most of the $z\leq9$ cases. In contrast, the (AGN-free) EDE case demonstrates more consistency at $z\sim11-12$ as well as the most luminous end of $z\leq10$. The corresponding $\chi^{2}$ results are displayed in Table \ref{tab:chi2}. Note that the values at $z=9$ are dominated by the outliers which only provide the upper limit of the LF. If they are removed, the other data fit better with the $\Lambda$CDM model. We speculate that there are three possible explanations for the outnumbered galaxy abundance at $z>9$ as follows,
\begin{enumerate}
    \item The pollution of the data that leads to errors in the luminosity measurements; \label{item:lf_1}
    \item The dust attenuation calibration method at that redshift range should be different from the lower redshifts; \label{item:lf_2}
    \item The efficient luminosity-sensitive feedback mechanisms that intrinsically suppress or partly quench the galaxy formation over redshift and luminosity, so that the luminosity function does not grow as fast as the EDE model expects. \label{item:lf_3}
\end{enumerate}

The calibration of the pollution is one of the major challenges of high redshift galaxy observations. According to Equation (\ref{eq:mM_relation}), there are in principle three possible causes that lead to the inaccuracy of the luminosity: the disturbance on the flux that pollutes the apparent magnitude $m$, the instrumental error of the telescope itself, and the uncertainty of the redshift $z$, Among them, the most commonly discussed error source is the uncertainty of the redshift from JWST. By the time our analysis was finished, the spectroscopic redshift $z_{\text{spec}}$ of the JWST galaxies were not yet available, while there were some reservations over the accuracy of their photometric redshift $z_{\text{photo}}$, e.g., \cite{2023arXiv230315431A}. However, considering the luminosity function contains the information of multiple galaxies, rather than the single. The chance is low that all of their $z_{\text{photo}}$ are systematically overestimated. The pollution from e.g., the selection effect or observational bias, on the other hand, is possible, as JWST brings unprecedented resolution of the low surface brightness objects. The detailed discussion on this matter \ref{item:lf_1} and an elaborate study on the high redshift dust attenuation calibration method \ref{item:lf_2}, however, are beyond the scope of this paper.

In this paper, we are most interested in the possibility \ref{item:lf_3}. Should the data sets themselves be accurate, and the dust attenuation not significantly change at higher redshifts, we argue that if the dark matter-dominated early Universe is better described by the EDE model, it needs to be accompanied by a baryonic feedback mechanism that considerably affects the galaxy formation process and suppress the growth of the galaxy abundance. Interestingly, as is shown in Figure \ref{fig:lf_data} and previous studies of the galaxy abundance at $z\leq10$, e.g., \cite{2020MNRAS.493.2059B,2023arXiv231114804C}, the deviating of the observed galaxy number density away from $\Lambda$CDM takes place at the luminous/massive end, and gradually fall back as the redshift decreases. This phenomenon suggests that the suppression mechanism we speculate should first affect the low-mass galaxies and transmit its effect to the massive end over the evolution of the galaxy abundance. Consequently, the pure dark matter-dominated EDE Universe at $z\sim10-12$ is transformed into the baryon/dark matter-dominated $\Lambda$CDM Universe at lower redshifts.

The comparison of the predicted and observed luminosity function, however, is not as direct as it looks. As mentioned above, the prediction of the luminosity function strongly depends on the stellar mass function, for there is no one-to-one bond between the halo mass and the luminosity of the corresponding galaxy. Consequently, although the scaling relation of $M_{\ast}-M_{\text{UV}}$ can correctly convert the stellar mass and the luminosity data upon which the scaling relation is built, it may not reliably reflect the properties of the data from other observations, for the estimation of the stellar mass by itself varies with the hypotheses of the Initial Mass Function (IMF), SFR and other critical conditions each work applies. As is shown in Figure 16 in \cite{2021ApJ...922...29S}, the scaling relations built upon various works can still be noticeably diverse. Be that as it may, now that we see the main clash between EDE prediction and LF observation lies on $z<10$, it becomes necessary to investigate further and compare our prediction with a recent result of the JWST galaxy abundance at lower redshift in terms of $M_{\ast}$, which fermented a ``turmoil'' on our previous understanding of the galaxy number density based on the pre-JWST observations.

\cite{2023Natur.616..266L} estimated the stellar mass of six massive galaxies at redshift $z\sim7.4-9.1$ from the early release of the JWST data and found that the stellar mass density, the integral of the stellar mass function above a given mass $M_{\ast}$, is about an order of magnitude larger than the extrapolation of the previous best-fitting from HST+Spitzer. As is mentioned in \cite{1999ApJ...521...64M}, the high redshift galaxies are treated as the scaled-up starburst galaxies. Hence it is common to estimate the mass of the high redshift galaxies by the locally confirmed IMF and other assumptions. \cite{2023NatAs...7..731B} found that according to the current understanding of the galaxy formation theory, the unexpectedly high abundance of these massive galaxies would require more than 50 per cent or even all of the baryons to form stars, which lies on the very edge of what is allowed by $\Lambda$CDM, and way beyond the empirical limit of the star formation efficiency shown in Figure \ref{fig:mhms_ratio}, thus put the current theory of the galaxy formation at the early time under the $\Lambda$CDM model into challenge.

\begin{figure}
    \centering
    \includegraphics[width=\columnwidth]{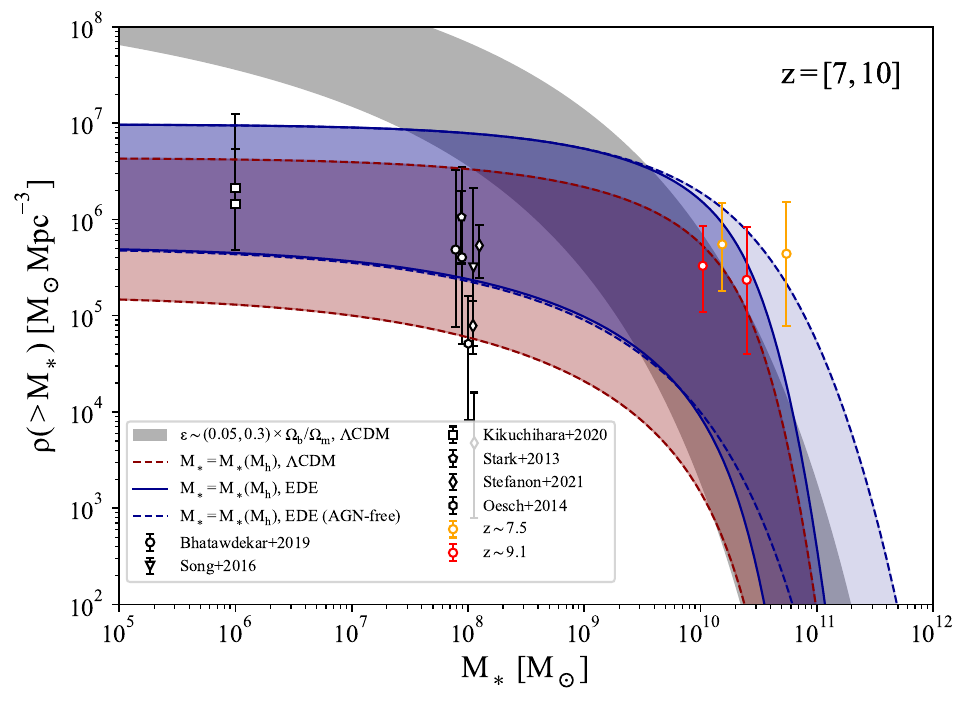}
    \caption{The stellar mass density under EDE (blue) and $\Lambda$CDM (red) cosmology. The blue dashed lines are the EDE with AGN-free scaling relation as is shown in Figure \ref{fig:mhms_ratio}. The blue and red shades span the redshift from $7$ to $10$ to match the range of the corresponding data. The grey shade area corresponds to the fixed $5-30$ per cent star formation efficiency at $z=7.5$. The black data are from the pre-JWST observations from \citet{2013ApJ...763..129S,2021ApJ...922...29S,2020ApJ...893...60K,2014ApJ...786..108O,2016ApJ...825....5S,2019MNRAS.486.3805B} and the coloured data are what \citet{2023Natur.616..266L} estimated from JWST. The pre-JWST data is consistent with the $\Lambda$CDM scenario, while the prediction under EDE, particularly the AGN-free case, matches the JWST results better.}
    \label{fig:smd}
\end{figure}

Should the pipeline to approach the stellar mass of the JWST galaxy candidates be appropriate and the cosmology be allowed to vary, we find that this phenomenon can be reproduced by the EDE model. In Figure \ref{fig:smd}, we demonstrate the stellar mass density that spans the range of $z\sim7-10$ in both EDE (blue) and $\Lambda$CDM (red) models. On top of that, considering the possible AGN-free scaling relation shown in Figure \ref{fig:mhms_ratio}, we also plot the stellar mass density without the feedback at the massive end (blue dashed). The JWST data are marked by orange ($z\sim7.5$) and red ($z\sim9.1$) points, and the black data points are the pre-JWST data observed by \cite{2013ApJ...763..129S,2021ApJ...922...29S,2020ApJ...893...60K,2014ApJ...786..108O,2016ApJ...825....5S,2019MNRAS.486.3805B}. It is clearly shown that the pre-JWST data are covered by the $\Lambda$CDM region, while the EDE region, particularly the AGN-free case, covers the JWST data better. This phenomenon might be related to the mass threshold of the data, i.e., the JWST data are integrated within a higher mass range than the pre-JWST ones. As is indicated in Figure \ref{fig:lf_data}, only the data at the luminous end display consistency with the EDE model at this redshift range. Therefore, if the suppression mechanism we speculate is able to explain all LF data, it should be expected to simultaneously work for all SMD data as well. A detailed study of this matter will be conducted for future work. In comparison, we also plot the $\Lambda$CDM case with $z=7.5$ in the grey shade area, assuming $\epsilon=(0.05-0.3)\times \Omega_{b}/\Omega_{m}$. In this scenario, however, neither pre-JWST nor JWST data points can be nicely explained. The reason is that the star formation efficiency of the massive galaxy-forming regions is normally higher, while the low-mass halos normally share a lower value. As a consequence, the empirical $M_{h}-M_{\ast}$ in Figure \ref{fig:mhms_ratio} determines the flattened and sharpened shape of the low mass and high mass region, respectively, in the stellar mass density shown in Figure \ref{fig:smd}.

Through the comparison between the theoretical prediction of the stellar mass function (Figure \ref{fig:smf}), luminosity function (Figure \ref{fig:lf_data}) and stellar mass density (Figure \ref{fig:smd}), we argue that the galaxy abundance at high redshift ($z\geq8$) indeed shows a promising prospect to verify the EDE model (or other models, as long as they can put a similar effect on the matter power spectrum), for it demonstrates considerable deviation under the EDE and $\Lambda$CDM model. Generally speaking, $z\sim11-12$ and the most luminous/massive data at lower redshifts show better consistency with EDE while the majority of the data at $z\leq9$ fit better with the $\Lambda$CDM case. The $z\sim10$ appears to be the transition era of the preference shift. This phenomenon suggests that neither $\Lambda$CDM nor EDE (the sole enhancement of the matter power spectrum, \cite{2024PhRvL.132f1002S}) alone accesses the explanation of the luminosity function evolution across this redshift range. Assuming the EDE model is a better description of the early Universe, a possible resolution to this phenomenon is to introduce a luminosity/mass-sensitive suppression mechanism that slows down or quenches the growth of the galaxy, beginning at the low mass halos and transmitting its effect to the massive ones as the Universe evolves to the lower redshift.

The intrinsic concern over the stellar mass density measured by JWST is that the scaling relation between luminosity and stellar mass at high redshift may not follow the same empirical rule constructed from the previous observations, e.g., the IMF might be different in the young galaxies in the early Universe compared to that in the local starbursts, which results in an overestimation of the stellar mass. The result of our calculation, however, provides an alternative interpretation that the abundance of the most massive galaxies could indeed be higher than previously expected because of the EDE model. Consequently, both the luminosity function and stellar mass density under this model are elevated and fit better with the high redshift and massive JWST data. The scaling relation, on the other hand, remains unchanged for the scale of the galaxy formation is too small to be strongly affected by the presence of EDE. Bearing all the discussions above in mind, albeit it is yet too soon to claim that another observational evidence of EDE has been found, due to the clash between the LF data and EDE at $z<10$, we argue that it is safe to regard the high redshift abundance of the galaxies as another probe of the EDE model. We also suggest that further investigations should focus on three aspects: 1. The detailed re-evaluation of the stellar mass under the EDE model; 2. More luminosity function data with $z\geq7$; 3. The possible luminosity/mass-sensitive suppression mechanism of the galaxy formation during the epoch of reionisation.

\section{Conclusions}
\label{sec:conclusions}

The early dark energy model is one of the few theories to resolve the Hubble tension that carries both theoretical capability and observational evidence. At present, the signal of EDE has been detected in some of the CMB observations. However, the verification of EDE suffers severe challenges on the small-scale (e.g., full-scale Planck CMB) and low-redshift (e.g., BOSS matter power spectrum) observations. Also, applying the EDE-like mechanism to resolve the Hubble tension will inevitably exacerbate the $S_{8}$ tension, for it incubates a denser Universe than the $\Lambda$CDM case, which already showed discordance with the weak lensing observational results. Considering the complications brought by the baryonic matter in the observations of the low redshift galaxies, we instead attempted to inspect if the EDE model can be probed by the high redshift galaxy abundance, for it provides an alternative scope to detect the small-scale density evolution of the Universe when it was less affected by the baryons.

Our investigation found that despite the complications of the galaxy formation processes, the current number density of the observed high redshift galaxies expressed in terms of the luminosity function, stellar mass function, and stellar mass density, detect some traces of evidence that prefer EDE. This is particularly true for the JWST data, as it discovered a surprisingly high number density of the luminous/massive galaxies which cannot be naturally explained by the $\Lambda$CDM model. Although the current data is not enough to put an unarguable Aye or Nay to the EDE model as yet, we argue that the galaxy abundance is indeed an efficient probe for its verification, and the upcoming JWST luminosity function data within redshift $z\sim8-12$, as well as the corresponding stellar mass estimation, can be expected to verify EDE independent of CMB and low redshift galaxy survey.

The way we applied in this paper to predict the galaxy abundance depends on the theoretical prediction of the halo mass function and the empirical scaling relations between $(M_{h}, M_{\ast}, M_{\text{UV}})$. Assuming the properties of the dark matter are unaffected by EDE, we found that the EDE and $\Lambda$CDM halo mass functions are almost identical at lower redshifts while significantly distinguishable when $z>4$, which is consistent with \cite{2021MNRAS.504..769K} (see Figure \ref{fig:hmf_comp}). This result is ideal for our study because the abundance matching method indicates that the corresponding observable stellar mass or luminosity function should share a similar detectability. 

Next, we estimated the corresponding stellar mass function under the two models. As shown in Figure \ref{fig:smf} and Table \ref{tab:chi2}, the existing pre-JWST data show a slight preference towards EDE over $\Lambda$CDM up to $z\sim10$. However, the preference highly depends on the choice of the data sources. The redshift-free scaling relation between $M_{h}$ and $M_{\ast}$ (Equation \ref{eq:mhmast}) indicates that the low mass halos are extremely inefficient in forming stars, while the turnover at $M_{h}\sim10^{11.5} \ M_{\odot}$ may need to be treated with extra care. At lower redshift, the turning point of the massive galaxy number density is normally caused by the AGN feedback, while the situation at higher redshift could be different due to the rareness of AGN in these extremely young galaxies. Considering the data to form the scaling relation in \cite{2021ApJ...922...29S} does not constrain the turning point at redshift $z>8$, it is reasonable to assume that $z\sim8$ is too early for most AGN to form. Therefore, we also extrapolated the scaling relation by making $\gamma=0.01$ rather than the original fit $0.4$ in Equation \ref{eq:mhmast} to flatten the turnover, such that more massive galaxies can form due to the lack of AGN feedback (Figure \ref{fig:mhms_ratio} and \ref{fig:smf}). The accurate fitting of the scaling relation, however, depends on whether the future high redshift observations that cover the stellar mass $M_{\ast}\sim10^{11.5} \ M_{\odot}$ detect a pervasive presence of AGN in the young galaxies and the turning point in the scaling relation.

On the contrary, the luminosity function demonstrates a clear distinction between $\Lambda$CDM and the EDE model. The prediction of it depends on the stellar mass function, the scaling relation between $M_{\ast}$ and $M_{\text{UV}}$, and the dust attenuation calibration rule. We applied the scaling relation obtained by \cite{2021ApJ...922...29S} for the consistency with the stellar mass function and the upper limit of the redshift. The dust attenuation calibration was adapted from the recent JWST observation. In addition, the change in cosmology also affects the distance modulus and consequently, the observational result of $M_{\text{UV}}$. Our calculation showed that EDE would cause a slightly fainter result by $\sim0.15$ mag. Finally, Our result in Figure \ref{fig:lf_data} and Table \ref{tab:chi2} suggests that the $\Lambda$CDM model is favoured by most of the observational results at $z\leq9$, while the $z\sim11-12$ as well as the most luminous data at lower redshifts show a preference for EDE. The $z\sim10$ appears to be a transition era of the two models. The contradiction between LF and SMF at $z\leq9$ might be related to the choice of data sources or the details in the estimation of $M_\ast$ and its variation under the EDE model, which is beyond the topic of this work.  Given that $z\sim10$ is indeed the boundary of this preference shift, we speculate that if EDE is to work for the entire evolution history of the galaxy abundance, it must be combined with a luminosity/mass-sensitive suppression mechanism that first affects the low mass end and gradually transmits its effect to the massive end during the epoch of reionisation, such that the early galaxy formation rate that follows the EDE model is slowed down and converted into the $\Lambda$CDM Universe.

The stellar mass density shown in Figure \ref{fig:smd} also implies a slight preference towards EDE. Akin towards SMF, the preference is sensitive to the choice of data sources. We found that the denser Universe produced by EDE can fit the JWST result observed by \cite{2023Natur.616..266L} better than the $\Lambda$CDM case. In particular, the AGN-free SHMR relation can further promote the consistency between the EDE prediction and the JWST data. Also, considering the EDE preference is shown in both the most luminous LF data and the SMD data with higher mass integration domain at $z\lesssim10$, we argue that the suppression mechanism speculated above needs to be able to explain all LF and SMD data simultaneously. However, the estimation of the stellar mass itself may also be affected by EDE via, e.g., the lower luminosity or a more efficient gas cooling process. In our calculation, we apply the vanilla results of the stellar mass since the impacts that the EDE model may inflict on the stellar mass estimation will (partly) cancel out each other. Nevertheless, a detailed evaluation of it is recommended for a more solid argumentation.

To conclude, our estimation of the galaxy abundance at $z\sim8-12$ suggests that it is capable of probing the $\Lambda$CDM and EDE model, for their difference can reach as much as an order of magnitude. However, neither model can fully depict the entire formation and evolution of the galaxy abundance around that epoch. For the $\Lambda$CDM model, an unrealistically high galaxy formation efficiency is required to explain the unexpectedly high number density of the galaxy at $z>10$ as well as the most luminous end at the lower redshifts. Equivalently, a successful EDE-based explanation of all data could work only if a luminosity/mass-sensitive suppression mechanism is confirmed. However, should it come to pass that the studies of days to come accord with our speculation, we might find it needful to contemplate the possibility that EDE be a fairer model to portray the early stage of the Universe, in the stead of $\Lambda$CDM.


\section*{Acknowledgements}


LW would like to thank Mauro Stefanon and Jiang, Junqian for their constructive and timely suggestions, and Roland Timmerman for organising the talk in Leiden where this work was first presented. LW, ZH and GY are supported by the National Key R\&D Program of China grant No. 2022YFF0503400, 2022YFF0503404 and China Manned Space Project grant No. CMS-CSST-2021-B01. GY is also supported by the National Key R\&D Program of China grant No. 2020SKA0110402, CAS Project for Young Scientists in Basic Research grant No. YSBR-092 and China Manned Space Project grant No. CMS-CSST-2021-A01. WX is supported by the National Natural Science Foundation of China grant No. 12373009, the CAS Project for Young Scientists in Basic Research grant No. YSBR-062, the science research grant from the China Manned Space Project, the Fundamental Research Funds for the Central Universities, and the Xiaomi Young Talents Program.
\section*{Data Availability}

The data underlying this article are available from published sources. The code to reproduce the results will be shared upon reasonable request.


\bibliographystyle{mnras}
\bibliography{ref} 

\bsp	
\label{lastpage}
\end{document}